# Studying focal ratio degradation of optical fibers for Subaru's Prime Focus Spectrograph

Jesulino Bispo dos Santos[1], Antonio Cesar de Oliveira[1], James Gunn[2], Ligia Souza de Oliveira[3], Marcio Vital de Arruda[1], Bruno Castilho[1], Clemens Darvin Gneiding[1], Flavio Felipe Ribeiro[1], Graham Murray[4], Daniel J. Reiley[5], Laerte Sodré Junior[6], Claudia Mendes de Oliveira[6]

1- MCT/LNA - Laboratório Nacional de Astrofísica, Itajubá - MG – Brazil
2- Department of Astrophysical Sciences - Princeton University - USA
3- OIO - Oliveira Instrumentação Óptica Ltda. – SP – Brazil
4- Centre for Advanced Instrumentation, Durham University, Physics Dept.-Durham – UK
5- Caltech Optical Observatories – Pasadena – CA - USA
6- IAG/USP – Instituto de Astronomia, Geofísica e Ciências Atmosféricas/ Universidade de São Paulo – SP - Brazil

**ABSTRACT**

Focal Ration Degradation (FRD) is a change in light's angular distribution caused by fiber optics. FRD is important to fiber-fed, spectroscopic astronomical systems because it can cause loss of signal, degradation in spectral resolution, or increased complexity in spectrograph design. Laboratório Nacional de Astrofísica (LNA) has developed a system that can accurately and precisely measures FRD, using an absolute method that can also measure fiber throughput. This paper describes the metrology system and shows measurements of Polymicro's fiber FBP129168190, FBP127165190 and Fujikura fiber 128170190. Although the FRD of the two fibers are low and similar to one another, it is very important to know the exact characteristics of these fibers since both will be used in the construction of FOCCoS (Fiber Optical Cable and Connectors System) for PFS (Prime Focus Spectrograph) to be installed at the Subaru telescope.

## 1.INTRODUCTION

Focal ratio degradation (FRD) is the effect whereby optical fibers tend to degrade the incident f-ratio to faster output f-numbers. Flaws such as micro-irregularities in the fiber, stress in the fiber, or polishing errors at the fiber end faces cause this phenomenon. FRD degrades the resolution of the spectrographs, reduces the signal on the spectrographs' sensors, and complicates spectrograph calibration. For Subaru's Prime Focus Spectrograph system (PFS), FRD is required to be less than 5%. The fibers need to be evaluated, considering that parts of optical fiber cables and fibers themselves may be subject to stress state in the operational conditions.

FRD is a challenging topic for several reasons. First, the absence of standardized measurement methods hinders both the communication of requirements and the comparison of measurements from different groups. Next, measuring FRD requires measuring faint halos of light near to main spot [01], a task that requires special instrumentation. Third, FRD measurements can be contaminated by external factors, such as misalignment between the fiber end and the input beam optical axis. Finally, different systems designs can place different requirements on FRD. Two classes of methods are used to measure FRD. The first method, called the "Absolute method", measures the output beam energy distribution and compares it to the energy of the input beam; this method measures transmission efficiency as well as FRD. The relative method measures only the output energy distribution. In this paper, we describe a system using the Absolute method. Subaru PFS instrument uses three sections of fiber optics. The first, Cable C, is built into the focal plane; it uses Polymicro FBP 127165190 fiber and is around 7m long. The last, Cable A, is built into the spectrograph; it uses Polymicro FBP 129168190 fiber and is around 55m long. Cable B links Cable A and Cable C, and is routed along the telescope structure; it uses Fujikura S.128/170 BPI fiber and is around 55m long. All three fibers have NA=0.22. Samples of all three fibers were tested as in this study.

## 2. EXPERIMENTAL PROCEDURES

Most of the FRD measurements to date are relative in nature; that is, they assume that all light is transmitted at some lower limit of the output f-ratio. While relative measurements are quite adequate for many purposes, they can be misleading when assessing the fibers performance at relatively fast f-ratios (f/# < 3.0) or looking at small core fibers. The scheme for absolute measurement, originally designed by Barden, compares the light in the input beam to that emanating from the fiber by way of a simple 90 degree flip of two separate mirrors. We have adapted a method described in Barden (1981) to measure absolute efficiency [02]. In our experimental set-up, although we can change the focal ratio of the input fiber, our plan was to feed the fibers only at f/2.85, close to the output focal ratio of Subaru's Hyper-Supreme corrector, F/2.8. In general, the technique uses **a science-grade** CCD (with frame grabber) and employs the IRAF routines QPHOT and PPOFILE (Tody, 1993) for data reduction [03]. Also it is possible to use other software to obtain curves of Absolute Transmission including characteristics of focal ratio degradation.

All fibers were mounted in US Conec MTP ferrules. To minimize stress on the fiber length, they were placed within a flexible tube, called a furcation tube. EPOTEK 301-2 was applied to the fiber and tubing to cement it in place; a 24-hour room temperature cure was used to further minimize stress. To minimize effects from the fiber end faces, all surfaces were polished using a 4-step polishing process and inspected to be optically flat and free from scratches.

**2.1 Experimental Setup**

The experimental apparatus consists of three pieces: an illumination arm, a monitoring arm and a measurement arm, Fig.1. The illumination arm illuminates the test fiber with an input beam of known focal ratio. The monitoring arm monitors the beam's position on the fiber end. The measurement arm measures the input beam and output beam, so they can be compared for a data reduction process.

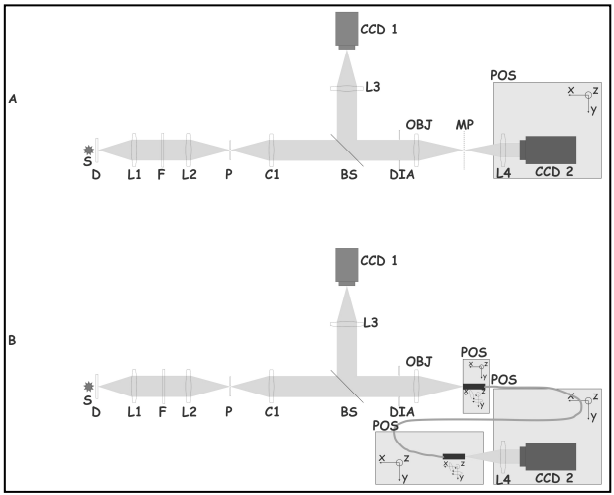
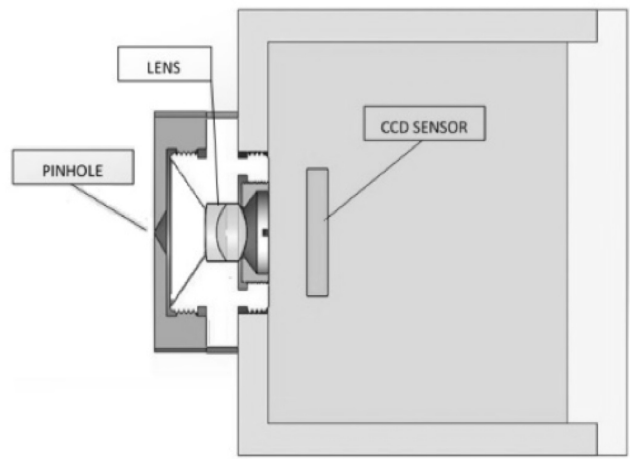

**Figure 1:** Diagram of the apparatus used to measure FRD in Absolute mode.

**Figure 2:** Main detector assembly – CCD2

The illumination arm consists of a halogen light source **S**, an Opal diffuser **D**, two lenses **L1** and **L2** (f=120mm), and a bandpass filter **F**. The light from **S** is diffused by **D1**, collimated by **L1** and focused by **L2** on the pinhole P (0.8mm diameter.) The light is then collimated by a long focal length lens C2 (f=400mm). The intended wavelength is selected placing the filter **F** between L1 and L2. The monitoring arm consists of a camera **CCD1** (Pixelink BP-741) positioned at the focal point of an achromatic doublet **L3** (f=400mm). It is connected to the measurement arm via beam splitter **BS**, and is used to monitor the position and alignment of the light beam on the fiber end. The measurement arm consists a diaphragm **DIA** placed in the front focal point of a Pentax camera lens **OBJ** (f=50mm), and a camera assembly placed on two micro-positioner stages **POS**. This optical arrangement is telecentric in the image space and produces on the measurement plane **MP** an f/2.85 beam focused in a 100um spot. The camera assembly, Fig. 2, consists of a pinhole (150µm), a lens **L4** (f =12 mm) and a camera **CCD2** (Andor Luca-S, 1002x1004, 8µm square pixel). The pinhole and the lens are fitted into a metal bracket, which can be fixed to the camera C-mount opening. To align the fibers with the light beam we use five degrees of freedom micro-positioner stages.

The instrument has two measurement settings. In setting **A**, the detector is moved to OBJ focal point and the P image is focused on aperture of the detector. The image grabbed in this setting (Reference Image) is used to determine the angular distribution of the light into the fiber. In setting **B**, one fiber end is placed on the measurement plane and the P image is focused on it. The other fiber end is positioned over the detector aperture. The image recorded by **CCD2** (Data Image) captures the angular distribution of the light out of the fiber; this image will be compared with the image obtained in configuration A in both intensity and spatial distribution.

## 2.2 Acquisitions and Data Reduction

We have developed a custom software package to reduce the fiber images and to obtain throughput energy curves. The package gives curves of enclosed energy. Fiber throughput is automatically determined as a function of output focal ratio. The first step is an estimation of the background level to be subtracted from the test exposures to remove the effects of hot pixels and stray light. The software then finds the image center by calculating the weighted average of all pixels. It associates a radius with each pixel and calculates the eccentricity; in the ideal case, this eccentricity is zero. Our target here is to obtain the absolute transmission of the fiber at a particular input f-ratio. To obtain the Absolute Transmission curve of the fiber at a particular Input Focal Ratio, Fig. 3, the software takes the concentric annulus centered on the fiber image.

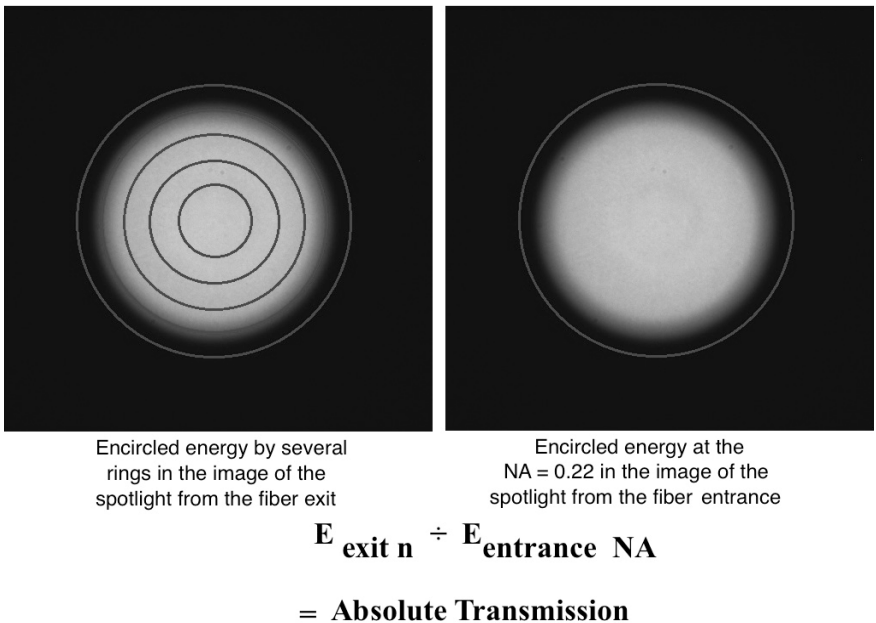

Encircled energy by several rings in the image of the spotlight from the fiber exit

Encircled energy at the NA = 0.22 in the image of the spotlight from the fiber entrance

$$E_{exit\ n} \div E_{entrance\ NA} = \text{Absolute Transmission}$$

**Figure 3:** Mathematic used to obtain the Absolute Transmission curves of the spotlight from the fiber exit and the light source at the fiber entrance.

This curve is used to define the efficiency over a range of f-numbers at the exit of the fiber, where each f-number value contains the summation of partial energy emergent from the fiber. Each energy value is calculated by the number of counts within each annulus divided by total number of counts from the reference image (setting A). The limiting focal ratio that can propagate in the tested fiber is approximately F/2.2 considering the numerical aperture of this fiber to be 0.22 ± 0.02. Therefore, we have defined NA=0.27 to be the outer limit (Region of integration) of the external annulus within which all of the light from the test fiber will be collected. The corresponding diameters of the annulus are converted to NA or Output Focal Ratios, taking in account camera pixel size of CCD2 and the lens **F4** focal length [04,05].

# 3. RESULTS AND CONCLUSIONS

## 3.1 Cable C - Polymicro fiber/ 7 meters

The optical fibers of each cable have slightly different core diameters such that the light is conducted from finer fibers for thicker fibers thereby improving the performance of connections. Cable C is in the telescope's focal plane, so it has the smallest core diameter. Fig. 4 shows plots of Absolute Transmission and Normalized Throughput for the Polymicro fiber FBP127165190, sample 7 meters long, in 3 specific wavelengths; 470 nm, 550 nm and 766 nm. Each curve is the average of 5 measurements, made over a one-hour interval. The curves for normalized throughput are almost identical for the three wavebands, indicating very negligible spectral variation in FRD. The Absolute Transmission curves saturate near the nominal NA for the fiber: NA= 0.22 ± 0.02. The value of this saturation agrees well with the specifications for the throughput for this fiber: 470 nm =>89%, 550 nm =>90.3% and 766 nm =>92%. This excellent agreement with fiber specifications builds confidence in the FRD measurements. FRD losses in PFS can be estimated using these FRD curves. For example, Fig. 5 shows the 550 nm curve from Figure 4, as well as the curve for the reference measurement and the curve for the theoretical case of no FRD. The vertical green line, NA = 0.2, represents the NA limit for light collected by the collimator in the PFS spectrograph. The intersection with the theoretical red curve shows that there is no loss from FRD in the nominal case. The intersection with the black, reference curve shows that this reference value also predicts negligible loss. The intersection with the blue, measured FRD, line shows that the predicted loss for this fiber around 1.3%. Fig. 6 shows a data frame from the fiber with the main rings used in the calculations.

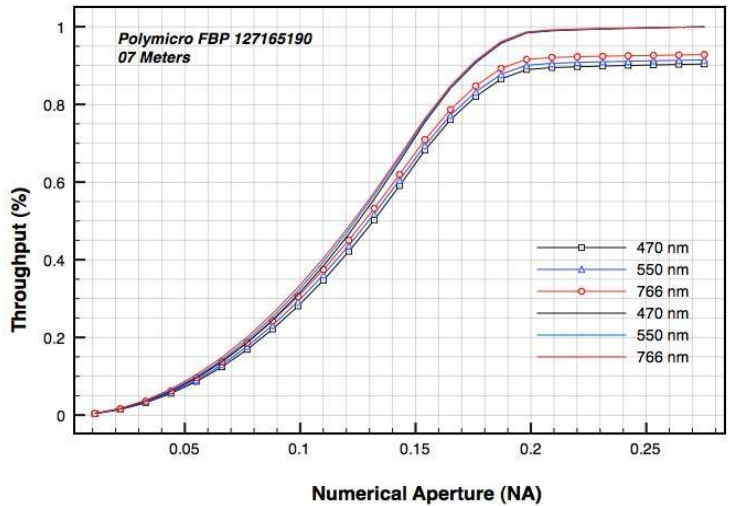

**Figure 4:** Plots showing Absolute Transmission curves, with symbols, and Normalized Throughput curves, without symbols, for Polymicro fiber 7-meters.

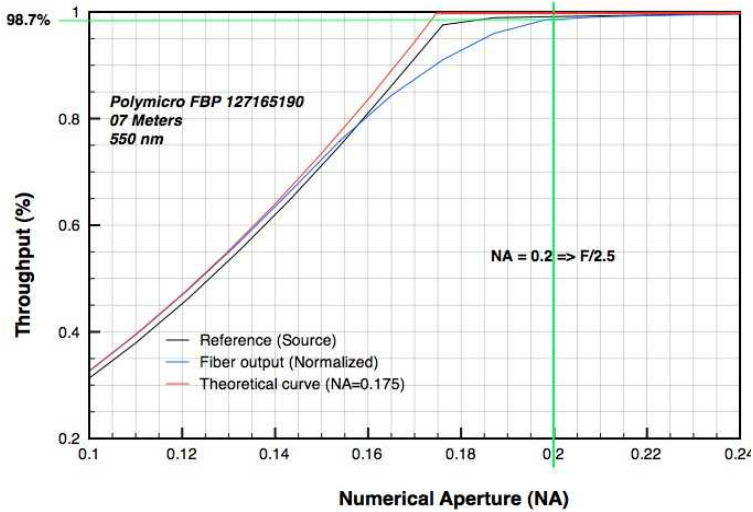

**Figure 5:** Throughput curves for Polymicro fiber FBP127165190, 7 meters, in 550 nm.

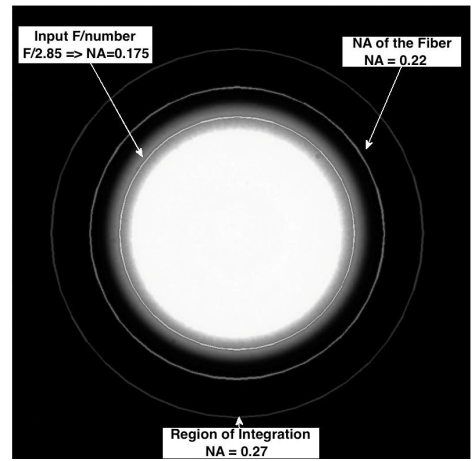

**Figure 6:** This image shows the focal ratio degradation beyond the F/2.85 annulus.

## 3.1 Cable B - Fujikura fiber/ 55 meters

Cable B receives light from Cable C, so it has slightly thicker fibers than the fibers from Cable C. Fig. 9, shows plots of Absolute Transmission and Normalized Throughput for the Fujikura S.128/170 BPI fiber, sample 55 meters long, at the same three wavelengths as before: 470 nm, 550 nm and 766 nm. Again, each curve is the average obtained from 5 measurements performed during one-hour interval. No spectral variation in FRD is observed at NA = 0.2. These curves also show excellent agreement with the manufacturer's specifications. NA= 0.22 ± 0.02 for; 470 nm =>72.5%, 550 nm =>80.7% and 766 nm =>88%. Figure 8 shows that FRD in 55 meters of this fiber will cause 4% loss in the PFS conditions of NA=0.175 input and NA=0.2 collected at the output.

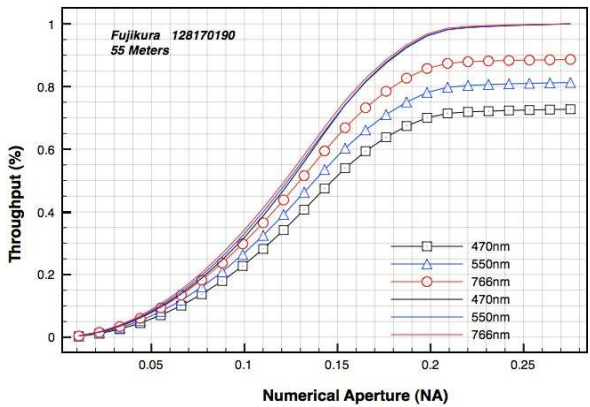

**Figure 7:** Plots showing Absolute Transmission curves, with symbols, and Normalized Throughput curves, without symbols, for Fujikura fiber 55-m.

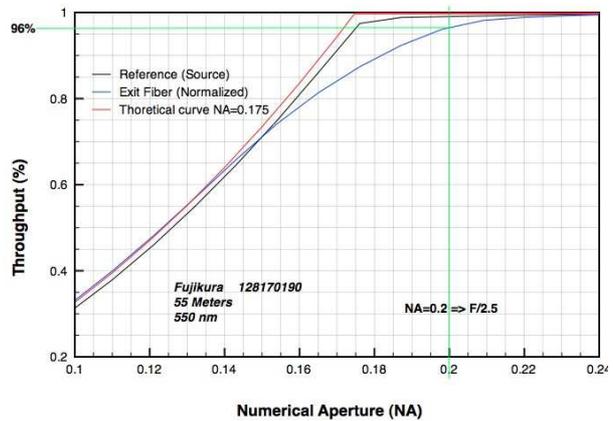

**Figure 8:** Throughput curves for Fujikura fiber, 55 meters, in 550 nm.

## 3.3 Cable A - Polymicro fiber/ 2 meters

Cable A receives light from Cable B, so it has the largest core diameter. Fig. 9, shows plots of Absolute Transmission and Normalized Throughput for the Polymicro fiber FBP129168190, 2 meters long, at the same three wavelengths as before: 470 nm, 550 nm and 766 nm. As with the other measurements, Fig. 10 shows excellent agreement with the manufacturer's specifications for both numerical aperture and throughput. Using the same methods described for the other fibers, Figure 12 shows FRD causes 2% loss for the conditions in PFS: NA=0.175 input and NA= 0.2 collected.

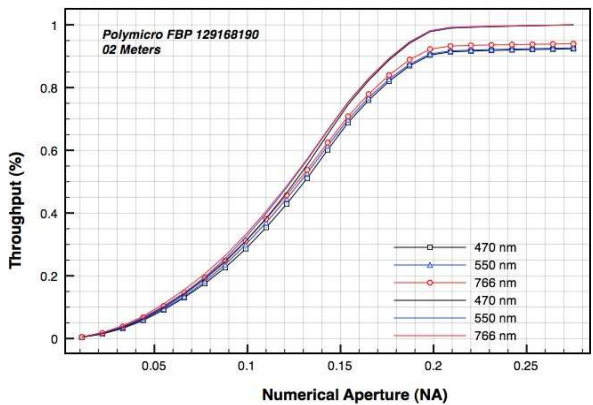

**Figure 9:** Plots showing Absolute Transmission curves, with symbols, and Normalized Throughput curves, without symbols, for Polymicro fiber 2-meters.

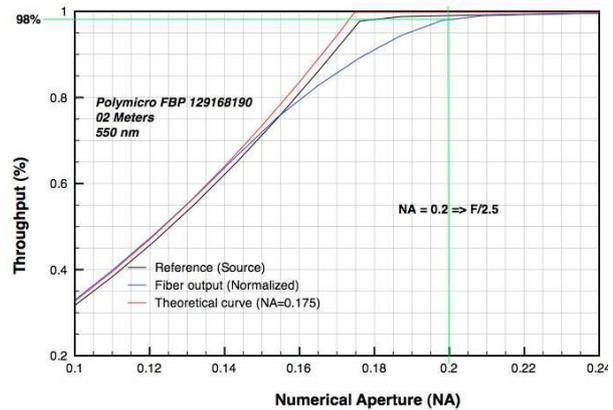

**Figure 10:** Throughput curves for Polymicro fiber FBP129168190, 2-meters, in 550 nm.

## 4. SUMMARY AND CONCLUSIONS

We presented in this paper a special study about the focal ratio degradation of optical fibers with a core size of 127µm, 128µm, and 129µm that will be used in the construction of FOCCoS (Fiber Optical Cable and Connectors System) for PFS (Prime Focus Spectrograph) /Subaru. All fibers had polyimide buffer. The 128-µm-core diameter fiber was manufactured by from Fujikura, and the other fibers were manufactured by Polymicro. The technique used in this work measures Absolute Transmission, comparing launched power to transmitted power, both as a function of numerical aperture. The normalized curves give us the FRD intrinsic to the test fiber, showing the throughput for each specific focal ratio. Analysis of the curves shows the mode dependent loss. These measurements of unstressed fiber provide a baseline for future measurements of FRD when the fiber is under the anticipated operational stresses. Table I shows a summary of the results of this work.

**Table I:** This table contains the relative efficiency from the Normalized Throughput and the Absolute Transmission for 3 different wavelengths, at input F/2.85 and output F/2.5

| Fiber Model | Normalized Throughput | Absolute Transmission | | |
|---|---|---|---|---|
| | 470 nm, 550 nm, 766 nm | 470 nm | 550 nm | 766 nm |
| Fujikura    S.128/170 BPI – (55 meters) | 96.8% | 70.5% | 78.5% | 86.1% |
| Polymicro FBP 127165190 – (07 meters) | 98.5% | 89.2% | 90.2% | 91.7% |
| Polymicro FBP 129168190 – (02 meters) | 98.2% | 90.6% | 90.8% | 92.5% |

## 5. ACKNOWLEDGMENTS

We gratefully acknowledge support from: The Funding Program for World-Leading Innovative R&D on Science and Technology. SUBARU Measurements of Images and Redshifts (SuMIRE), CSTP, Japan. INCT-Astrofísica (INCT-A/CNPq), Brasil. Fundação de Amparo a Pesquisa do Estado de São Paulo (FAPESP), Brasil. Laboratório Nacional de Astrofísica, (LNA) e Ministério da Ciência Tecnologia e Inovação, (MCTI), Brasil.